\documentstyle[twoside,fleqn,espcrc2,epsfig]{article}
\newcommand{\deu}{(\delta^u_{23})}

\newcommand{\bxs}{ B \rightarrow X_s l^+\ell^-}
\newcommand{\gcenu}{ B \rightarrow X_c e \nu}
\newcommand{\ds}{\displaystyle}

\renewcommand{\d}{\delta}

\newcommand{\g}{\gamma}

\newcommand{\G}{\Gamma}

\newcommand{\bea}{\begin{eqnarray}}
\newcommand{\eea}{\end{eqnarray}}
\newcommand{\beq}{\begin{equation}}
\newcommand{\eeq}{\end{equation}}
\newcommand{\nn}{\nonumber}
\newcommand{\fr}{\frac}
\newcommand{\hl}{\hline}

\newcommand{\AmS}{{\protect\the\textfont2
  A\kern-.1667em\lower.5ex\hbox{M}\kern-.125emS}}

\title{\Large A model independent analysis of  $\mathbf B \rightarrow X_s \ell^+ \ell^-$
   decays in supersymmetry}

\author{I. Scimemi,
Dep. de Fisica Teorica, Univ. de Valencia,
c. Dr.Moliner 50, E-46100, Burjassot, Valencia, Spain
\thanks{I  acknowledge  E. Lunghi, A. Masiero, L. Silvestrini for  collaboration  in 
 the elaboration of  the work.
I thank Della Riccia Foundation (Florence, Italy) for support.
Preprint FTUV/99-77, IFIC/99-80}}

\begin{document}

\begin{abstract}
Recently  the semileptonic decays $B\rightarrow X_s e^+ e^-$, 
$B\rightarrow X_s \mu^+ \mu^-$ 
in generic supersymmetric extensions of the Standard Model have been studied in ref.~\cite{noi}.
In this talk I review the main points of this analysis.
SUSY effects are parameterized using the mass insertion approximation formalism
 and differences with MSSM results are pointed out. Constraints on SUSY
contributions coming from other processes ({\it e.g.} $b\rightarrow s \g$) are taken into
account. Chargino and gluino contributions to photon and Z-mediated decays are
computed and non-perturbative  corrections are considered.
We find that the integrated branching ratios  and the asymmetries can
be strongly modified.
Moreover, the behavior of the differential Forward-Backward asymmetry
remarkably changes with respect to the  Standard Model expectation. 

\end{abstract}
\maketitle

\section{Introduction}
\label{sec:intro}
One of the features of 
 a general low energy supersymmetric (SUSY) extension of  
 the Standard Model (SM) is
the presence of a huge number of new parameters. FCNC and CP violating
 phenomena constrain strongly a big part of the new parameter space. 
However there is still room for significant departures from the SM
 expectations in this interesting class of physical processes.

Recently we have  investigated the  relevance
 of new physics effects in the 
semileptonic inclusive decay $B \rightarrow X_s
\ell^+ \ell^-$~\cite{noi}.
This decay is quite suppressed in the Standard Model;
however, new $B$-factories should  reach the precision requested by the
 SM prediction~\cite{babar} and an estimate of all possible new contributions to
 this process is compelling.

Semileptonic charmless $B$ decays have been deeply studied.
The dominant perturbative SM contribution has been evaluated  in 
ref.~\cite{grin} and later two loop QCD corrections have been 
provided~\cite{bura,bura1}. The contribution due to  $c\bar c$ resonances 
 to these results are included
in the papers listed in ref.~\cite{tram}.
Long distance corrections   can  have a different  origin  according to
the value of the dilepton invariant mass one considers.
$O(1/m_b^2)$ corrections have been first calculated  
in ref.~\cite{falk} and recently  corrected in ref.~\cite{alih,buch}. 
Near the peaks, non-perturbative contributions generated by $c\bar c$ resonances  by means of
resonance-exchange models have been  provided in ref.~\cite{alih,krug}.
Far from the resonance region, instead, ref.~\cite{rey} (see also ref.~\cite{rupa}) 
estimate $c\bar c$ long-distance effects
using a heavy quark expansion in inverse powers of the
 charm-quark mass ($O(1/m_c^2)$ corrections).

An analysis  of the SUSY contributions
 has been presented in refs.~\cite{cho,goto} 
 where the authors estimate
 the contribution of the Minimal Supersymmetric
Standard Model (MSSM). 
They consider first  a universal soft supersymmetry breaking sector
at the Grand Unification scale (Constrained MSSM) 
and then partly relax this universality condition.
In the latter case they  find that there can be a substantial difference between
the SM and the SUSY results in the Branching Ratios and in the
 forward--backward asymmetries. One of the reasons of this enhancement
is that the Wilson coefficient $C_7(M_W) $  can change sign 
with respect to the SM in some region of the parameter space
 while respecting constraints coming from $b\rightarrow s \g$.
The recent measurements of   $b\rightarrow s \g$~\cite{cleo}
 have narrowed the window
of the possible values of  $C_7(M_W)$ and in particular a sign change of this
coefficient is no more allowed in the Constrained MSSM framework.
Hence, on one hand it is worthwhile considering $B \rightarrow X_s \ell^+
 \ell^-$ in a more general SUSY framework then just the Constrained MSSM, and, on the
 other hand, the above mentioned new results prompt us to a
 reconsideration of the process.
 In reference~\cite{korea} the possibility
 of new-physics effects coming from gluino-mediated FCNC is studied.

We consider all possible contributions to charmless semileptonic $B$ decays 
 coming from chargino-quark-squark and gluino-quark-squark interactions
and we analyze both Z-boson and photon mediated decays.
Contributions coming from penguin and box diagrams are taken into account;
moreover, corrections to the MIA results due to a light $\tilde t_R$
are considered.
A direct comparison between the SUSY and the SM contributions to
the Wilson coefficients  is performed.
Once   the constraints on mass insertions are established,
we find that in generic SUSY models
 there si still  enough room  in order to see large deviations from
 the SM expectations for branching ratios and asymmetries. 
For our final computation of physical observables
 we consider NLO order QCD evolution of the coefficients and 
non-perturbative corrections ($O(1/{m_b^2}),\ O(1/{m_c^2}),$..),   
each in its proper range of the dilepton invariant mass.

Because of the presence of so many unknown parameters (in particular in the
scalar mass matrices) which enter in a quite complicated way in
the determination of the mass eigenstates and of the various mixing matrices
it is very useful to adopt the so-called 
``Mass Insertion Approximation''(MIA)~\cite{hall}. 
In this framework one chooses a basis for fermion and sfermion states
in which all the couplings of these particles to neutral gauginos are
 flavor diagonal. Flavor changes in the squark sector are provided
 by the non-diagonality of the  sfermion propagators.
The pattern of flavor change is then given by the ratios 
\beq
(\delta^{f}_{ij})_{AB}= 
\fr{(m^{\tilde f }_{ij})^2_{AB}}{M_{sq}^2} \ ,
\eeq
where $ (m^{\tilde f }_{ij})^2_{AB}$  are the off-diagonal elements of the
$\tilde f=\tilde u,\tilde d $ mass squared matrix that
 mixes flavor $i$, $j$ for both left- and
right-handed scalars ($A,B=$Left, Right) and  $M_{sq}$ is the
average squark mass (see {\it e.g.}~\cite{gabb}).
The sfermion propagators are expanded in terms of the $\delta$s
and the contribution of the first two terms of this expansion are considered.
We  show that the  graphs with a double MI can be safely neglected in this process.
The genuine SUSY contributions to the Wilson coefficients will be simply
proportional to the various $\delta$s and a keen analysis of 
the different Feynman diagrams involved will allow us to isolate the few
insertions really relevant for a given process. 
In this way we see that only a small number of the new parameters is
involved and a general SUSY analysis is made possible.
The hypothesis regarding the smallness of the $\delta$s and so the
reliability of the approximation can then be checked {\it a posteriori}.

Many of these $\delta$s are strongly constrained
by FCNC effects~\cite{gabb,hage,gab} or by
vacuum stability arguments~\cite{casa}.
Nevertheless  it may happen  that such limits are not strong enough to prevent
large contributions to some rare processes.

\section{Operator basis and general framework}
\label{sec:opba}

 The effective Hamiltonian for the decay $ B \rightarrow X_s \ell^+\ell^-$
in general  low-energy SUSY models
  is  the same in the SM~\cite{bura,bura1} and
in the MSSM~\cite{cho,goto}, it  is known at next-to-leading order
 and we refer  to  the cited articles for its expression.
We  find that  SUSY  can
also modify (with respect to the SM) the  matching
coefficients of the operators
 \bea
Q_7^\prime&=&\fr{e}{8 \pi^2}m_b\bar{s}_R\sigma^{\mu\nu}b_L F_{\mu\nu},\nn\\
Q_9^\prime&=&(\bar{s}_R\g_\mu b_R)\bar{l}\g^\mu l, \nn \\
Q_{10}^\prime&=&(\bar{s}_R\g_\mu b_R)\bar{l}\g^\mu \g_5 l .
\label{eq:ope}
\eea
However   we have checked  that the contribution
of these operators is negligible and so they are not
considered in the final discussion of physical quantities.
SUSY contributions to other operators are of higher perturbative order and can be neglected.

The observables we  have in mind  are the differential branching ratio and 
the forward-backward asymmetry,
\begin{eqnarray}
R(s) &\equiv& {{\rm d} \ \G (\bxs) / {\rm d}  s\over \G (\gcenu)}  
\label{eq:br0} \\
A_{FB}(s) & \equiv & \nn 
\end{eqnarray}
\beq
 {\ds \int_{-1}^1 {\rm d} \cos{\theta} \; {\ds{\rm d}^2 
		     \G (\bxs)\over
                     \ds {\rm d} \cos{\theta} \; {\rm d} s} \; {\rm Sgn} (\cos{\theta})    
                     \over
                     \ds \int_{-1}^1 {\rm d}\cos{\theta} \; {\ds {\rm d}^2 \G (\bxs)\over
                     \ds {\rm d}\cos{\theta} \; {\rm d} s}}  
\label{eq:afb0} 
\eeq
where $s=(p_{\ell^+}+p_{\ell^-})^2 /m_b^2$, $\theta$ is
the angle between the positively charged lepton and the B flight
direction in the rest frame of the dilepton system.

It is worth underlying that integrating the differential asymmetry
given in eq.~(\ref{eq:afb0}) we do not obtain the global
Foward--Backward asymmetry which is by definition:
\bea
{N(\ell^+_\rightarrow) - N(\ell^+_\leftarrow)
           \over N(\ell^+_\rightarrow) + N(\ell^+_\leftarrow)} &\equiv&
\nn 
\eea
\beq
 {\ds \int_{-1}^1 {\rm d} \cos{\theta} \int {\rm d} s\; {\ds{\rm d}^2 
		  \G (\bxs)\over
                     \ds {\rm d} \cos{\theta} \; {\rm d} s} \; {\rm Sgn} (\cos{\theta})    
                     \over
           \ds \int_{-1}^1 {\rm d}\cos{\theta} \int{\rm d} s \; {\ds {\rm d}^2 \G (\bxs)\over
                     \ds {\rm d}\cos{\theta} \; {\rm d} s}} 
\label{intafb2}
\end{equation}
where $\ell^+_\rightarrow$ and $\ell^+_\leftarrow$ stand respectevely for
leptons scattered in the forward and  backward direction.

\noindent To this extent it is useful to introduce the following quantity
\bea
\overline A_{FB} (s)  &\equiv & 
\eea
\beq
 {\ds \int_{-1}^1 {\rm d} \cos{\theta} \; {\ds{\rm d}^2 
		     \G (\bxs)\over
                     \ds {\rm d} \cos{\theta} \; {\rm d} s} \; {\rm Sgn} (\cos{\theta})    
                     \over
           \ds \int_{-1}^1 {\rm d}\cos{\theta} \int {\rm d} s \; {\ds {\rm d}^2 \G (\bxs)\over
                     \ds {\rm d}\cos{\theta} \; {\rm d} s}}   
\label{eq:afb20} 
\eeq
 whose integrated value is given by eq.~(\ref{intafb2}).

Eqns.~(\ref{eq:br0}) and (\ref{eq:afb0}) have been corrected in order
 to include several non-perturbative effects.
We refer to~\cite{noi} and to references therein 
for all the definitions concerning this issue.  

\section{Light $\tilde t_R$ effects}
\label{sec:lightstop}
In the Mass Insertion Approximation framework we assume that all the
diagonal entries of the scalar mass matrices are degenerate and that
the off diagonal ones are sufficiently small. In this context we
expect all the squark masses to lie in a small region around an average
mass which we have chosen not smaller than 250 GeV. Actually there is
the possibility for the $\tilde t_R$ to be much lighter; in fact the
lower bound on its mass is about 70 GeV.
For this reason it is natural to wonder how good is the MIA when a
$\tilde t_R$ explicitly runs in a loop. 

The diagrams, among those we have computed, interested in this effect are 
the chargino penguins and box with the $(\delta^u_{23})_{LR}$
insertion. To compute the light--$\tilde t_R$ contribution we adopt
the approach presented in ref.~\cite{luca}. There the authors 
consider an expansion valid for unequal diagonal entries which
gives exactly the MIA in the limit of complete degeneration.  

\section{Constraints on mass insertions}
\label{sec:deltas}

In order to establish how large 
the SUSY contribution to $B\rightarrow X_s \ell^+ \ell^-$ can be, 
one can compare, coefficient
per coefficient, the MI results with the SM ones 
 taking into account possible constraints
on the $\d$s coming from other processes, in particular from $b\rightarrow s \g$.
A discussion about this issue can be  found in ref.~\cite{noi}.
The most relevant $\d$s interested in the determination of the Wilson
coefficients $C_7$, $C_9$ and $C_{10}$ are
$(\d_{23}^u)_{LL}$, $(\d_{23}^u)_{LR}$, $(\d_{33}^u)_{RL}$,
$(\d_{23}^d)_{LL}$ and  $(\d_{23}^d)_{LR}$.

\section{Results}
\label{sec:results}

While the gluino sector of the theory is essentially determined by the
knowledge of the gluino mass (i.e.~$M_{gl}$), the chargino one needs
two more parameters (i.e.~$M_2$, $\mu$ and $\tan \beta$). 
Moreover it is
 a general feature of the models we are studying the decoupling of
the SUSY contributions in the limit of high sparticle masses:
we expect the biggest SUSY contributions to appear for such masses
chosen at the lower bound of the experimentally allowed region.
On the other hand this considerations  suggest us to constrain the
 parameters of the chargino sector by the requirement of the lighter eigenstate not
to have a mass lower than  the experimental bound of  about 70 GeV~\cite{pdg}. 
 The  remaining two dimensional parameter space has  yet no  constraint. 
For these reasons we scan the chargino parameter space by means of
scatter plots~\cite{noi}.

Thus, with $\mu\simeq -160$, $M_{gl}\simeq M_{sq} \simeq 250$ GeV,
 $M_{\tilde \nu}\simeq 50$ GeV, $\tan \beta\simeq 2$
 one gets
\bea
C_9^{MI}(M_B) &=& -1.2 \deu_{LL} + 0.69 \deu_{LR} \nn \\
 & &  
 -0.51(\delta^d_{23})_{LL} \nn \\
      C_{10}^{MI}(M_B) &=& 1.75 \deu_{LL} - 8.25 \deu_{LR} 
\ .
\nn
\eea

In order to numerically compare  these values with  the respective SM 
results we note that the minimum value of
 $\left(C_9^{\rm eff}(s)\right)^{SM}(M_B)$ is about 4 
while $C_{10}^{SM} = -4.6$. 
Thus one   deduces that 
SM expectations  for the observables  are 
 enhanced when $C_9^{MI}(M_b)$ is positive.
 Moreover  the big value of   $C_{10}^{MI}(M_B)$  implies that
the final  total coefficient $C_{10}(M_B)$ can  have a different sign with 
respect to the SM estimate.
As a  consequence of this,
 the sign of   asymmetries can  be the opposite of the
one calculated in the SM.

The sign and the value of  the coefficient $C_7$ has a great
importance.
In fact the  integral  of the BR (see eq.~(\ref{eq:br0})) is dominated
 by the $|C_7|^2/s $  and  $C_7  C_9$ term  for low values of $s$.
In the SM the interference between $O_7$ and $O_9$ is destructive and
this behavior can be easily modifed in the general class of models we
are dealing with.
 It is worthwhile to  note  that the Constrained MSSM cannot drive a change in the the sign of $C_7$ while this can be realized in these kind of models.

\begin{table*}[hbt]
\setlength{\tabcolsep}{1.5pc}
\caption{\it Integrated BR, $A_{FB}$ and $\overline A_{FB}$ in
the SM and in a general SUSY extension of the SM for the decays
$B \rightarrow X_s e^+ e^-$ and $B \rightarrow X_s \mu^+ \mu^-$. 
The second and third columns
are the extremal values we obtain with a positive $C_7^{eff}$ while
the fourth one is the $C_7^{eff}<0$ case. 
The actual numerical inputs
for the various coefficients can be found in the text.
The BR is just the integral of $R(s)$ multiplied by the BR of the
semileptonic dominant B decay ($BR(B \rightarrow X_c e \nu) = 0.105$). }
\label{tab:resinteg}
\begin{tabular*}
{\textwidth}
{@{}l@{\extracolsep{\fill}}ccccc} 
\hline
Observable & SM & SUSY        &SUSY    &  SUSY    \\  
           &    & maximal     &minimal & ($C_7<0$) \\ \hline \hline
\vphantom{\fbox{$BR (e)$}}$BR (e)$&$9.6\,{{10}^{-6}}$&$4.3 \
10^{-5}$&$3.9\,{{10}^{-6}}$&$3.9 \ 10^{-5}$  \\ \hl 
\vphantom{\fbox{$BR (e)$}}$A_{FB} (e)$&$0.23$&$0.33$&$-0.18$&$0.31$  \\ \hl 
\vphantom{\fbox{$BR (e)$}}$\overline A_{FB} (e)$&$0.071$&$0.24$&$-0.19$&$0.11$  \\ \hline \hline 
\vphantom{\fbox{$BR (e)$}}$BR (\mu)$&$6.3\,{{10}^{-6}}$&$4.0 \
10^{-5}$&$1.6\,{{10}^{-6}}$&$3.4 \ 10^{-5}$  \\ \hl 
\vphantom{\fbox{$BR (e)$}}$A_{FB} (\mu)$&$0.23$&$0.33$&$-0.18$&$0.31$  \\ \hl 
\vphantom{\fbox{$BR (e)$}}$\overline A_{FB} (\mu)$&$0.11$&$0.27$&$-0.27$&$0.15$  \\ \hl 
\end{tabular*}
\end{table*}

The integrated BRs and asymmetries for the decays 
$B\rightarrow X_s e^+ e^-$ and 
$B\rightarrow X_s \mu^+ \mu^-$ in the SM case
and in the SUSY one (with the above choices of the parameters) are
summarized in tab.\ref{tab:resinteg}.
There we computed the total perturbative contributions neglecting the
resonances.

The results of tab.~\ref{tab:resinteg} must be compared with 
the experimental best limit which reads~\cite{cleodue}
\begin{eqnarray}
BR_{exp}&<& 5.8 \; 10^{-5}.
\end{eqnarray}

Looking table~\ref{tab:resinteg}
 we see                 
that the differences between SM and SUSY predictions can be remarkable.
 Moreover a sufficiently precise measure of  BRs, $A_{FB}$s and
$\overline A_{FB}$s
can either discriminate between the  CMSSM and
more general SUSY models or give new constraints on mass insertions.
Both these kind of informations can be very useful for model building. 

\section{Conclusions}
\label{sec:conclusions}
 In this paper a  discussion about  SUSY contributions to
 semileptonic decays $B\rightarrow X_s e^+ e^-$, 
$B\rightarrow X_s \mu^+ \mu^-$ 
 is provided.

Given the constraints coming from  the 
 recent  measure of $b\rightarrow s \g$
 and estimating all possible SUSY effects in the MIA framework 
 we see that SUSY   has  a  chance  to strongly
enhance or depress semileptonic charmless B-decays.
The  expected direct measure will give very interesting informations
about the SM and its possible extensions.

\end{document}